# The Magnitude and Frequency Variations of Vector-Borne Infections Outbreaks with the Ross-Macdonald Model: Explaining and Predicting Outbreaks of Dengue Fever

*Marcos Amaku[1], Franciane Azevedo[1], Marcelo Nascimento Burattini[1,2], Giovanini Evelim Coelho[3], Francisco Antonio Bezerra Coutinho[1], Luis Fernandez Lopez[1,4], Rogério Motitsuki[1], Annelies Wilder-Smith[5] and Eduardo Massad[1,6]\**

[1] LIM01-Hospital de Clínicas, Faculdade de Medicina Universidade de São Paulo, São Paulo, SP, Brazil

[2] Hospital São Paulo, Escola Paulista de Medicina, Universidade Federal de São Paulo, São Paulo, SP, Brazil

[3] Ministério da Saúde, Brasília, DF, Brazil

[4] Center for Internet Augmented Reserch & Assessment, Florida International University, Miami, FL, USA

[5] Lee Kong Chian School of Medicine, Nanyang University, Singapore

[6] London School of Hygiene and Tropical Medicine, London, UK

*with the exception of the corresponding author (edmassad@usp.br) the author are listed in alphabetical order.

**Running Title:** *Dengue Patterns of Propagation*

# Abstract

It is possible to model vector-borne infection using the classical Ross-Macdonald model. This attempt, however fails in several respects. First, using measured (or estimated) parameters, the model predicts a much greater number of cases than what is usually observed. Second, the model predicts a single huge outbreaks that is followed after decades of much smaller outbreaks. This is not what is observed. Usually towns or cities report a number of cases that recur for many years, even when environmental changes cannot explain the disappearance of the infection in-between the peaks. In this paper we continue to examine the pitfalls in modeling this class of infections, and explain that, in fact, if properly used, the Ross-Macdonald model works, can be used to understand the patterns of epidemics and even, to some extents, to make some predictions. We model several outbreaks of dengue fever and show that the variable pattern of year recurrence (or absence of it) can be understood and explained by a simple Ross-Macdonald model modified to take into account human movement across a range of neighborhoods within a city. In addition, we analyze the effect of seasonal variations in the parameters determining the number, longevity and biting behavior of mosquitoes. Based on the size of the first outbreak, we show that it is possible to estimate the proportion of the remaining susceptibles and predict the likelihood and magnitude of eventual subsequent outbreaks. The approach is exemplified by actual dengue outbreaks with different recurrence patterns from some Brazilian regions.

# Introduction

Dengue is a human disease caused by 4 related but distinct strains of a flavivirus and transmitted by urban vectors [1-3]. It is currently considered the most important vector-borne infection, affecting almost four hundred million people every year in tropical countries [4]. Its outbreaks recur with patterns of different magnitude and frequency that challenges our understanding with the current knowledge about the environment-mosquito-human interaction.

The Brazilian government maintains a data base of the weekly incidence of dengue for a large number of cities in all regions of the country. This data base shows a great variety of patterns as shown in the four examples in figure 1.

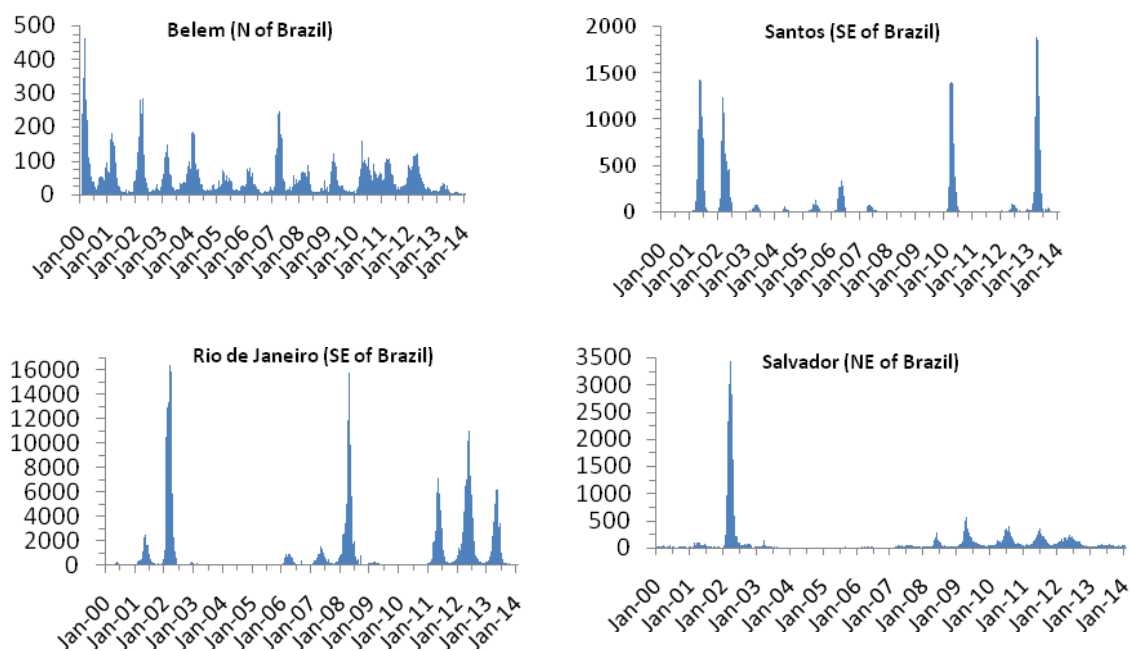

**Figure 1. Four different patterns of time-distribution of dengue outbreaks in Brazil (2000-2014)**

This paper has three aims:

1) To understand the origin and mechanism that produces so many different patterns;

2) To propose a mathematical model that aims to explain those patterns and to investigate if the mechanisms involved are unique.

3) To identify variables and parameters that serve to predict when the Dengue transmission ceases in a certain city and if not where it will occur in the future with a certain probability. These probabilities depend on certain measurements that may be demanding. So our third goal on this paper should be titled: How to forecast dengue epidemics.

This paper is organized as follows. In section 1 we describe a very simple model of dengue epidemics that incorporates elements that are new to usual dengue models. These elements were partially described in two previous papers by Amaku et al.[3], [5]. This model is in fact the classical Ross-Macdonald model but modified to avoid simple pitfalls in its application [5].

In section 2, we analyze the model and show that it can predict any pattern of the epidemics. We also exemplify the production of patterns with two mechanisms: movement of people and seasonality. In section 3, we introduce the calculation of a few quantities that should allow public authorities to make predictions about future events so that resources could be better allocated. Finally, in section 4, we present a summary of our findings.

## 1. The real vector of Dengue are the Humans

*Aedes aegyptii*, the dengue mosquito, has a very small life span, typically lasting for one week [3] and a short flight range (typically of one hundred meters). During this short life and flight-range span, therefore, the mosquito covers a very small area of the region where the dengue epidemics occur. Hence, considering that in big urban areas dengue circulates from one neighborhood to another, we are forced to admit that dengue is carried by humans that are infected but still able to move around.

Based on that, we propose a very simple model that consists in N regions of a city that are more or less geographically apart from each other, in the sense that the inhabitants of each of those regions rarely (or at the wrong time of the day) visit the other regions. By this we mean to visit the other regions at a certain time and for a certain period of time that would allow the humans being bitten by the mosquitoes and acquire the infection if one of them is infected. We will show that this explain why in some cities there can be a dengue epidemic in a district and not in another very near neighborhood in the same transmission season.

The model consists in the following system of differential equations. The meaning of the symbols are given in Table 1 to help the reader. $H_{Si}$, $H_{Ii}$ and $H_{Ri}$ are the densities of susceptible, infected and recovered individuals that live in the neighborhood *i*. $M_{Si}$, $M_{Li}$ and $M_{Ii}$ are the densities of susceptible, latent and infected mosquitoes that live in neighborhood *i*. The neighborhoods are assumed homogeneous and the populations are obtained by multiplying the densities by the area of each neighborhood.

| Table 1. Variables, parameters, their biological meaning and values of system (1) | | |
|---|---|---|
| **Variable** | **Biological Meaning** | **Initial Value** |
| $H_{Si}$ | Susceptible Humans | Variable |
| $H_{Ii}$ | Infected Humans | Variable |
| $H_{Ri}$ | Recovered Humans | Variable |
| $M_{Si}$ | Susceptible Mosquitoes | Variable |
| $M_{Li}$ | Latent Mosquitoes | Variable |
| $M_{Ii}$ | Infected Mosquitoes | Variable |
| **Parameter** | **Biological Meaning** | **Value** |
| $a_1$ | Biting rate of infected mosquitoes | 7 week$^{-1}$ |
| $a_2$ | Biting rate of non-infected mosquitoes | 7 week$^{-1}$ |
| $b$ | Probability of Transmission from Mosquitoes to Humans | 0.6 |
| $c$ | Probability of Transmission from Mosquitoes to Humans | 0.5 |
| $\beta_{ij}^H$ | Proportion of individuals from neighborhood $i$ that visit neighborhood $j$ | variable |
| $\mu_H$ | Mortality rate of Humans | 2.74 x 10$^{-4}$ week$^{-1}$ |
| $\Lambda_i^H$ | Growing rate of Humans | Variable (usually zero) |
| $\gamma_H$ | Humans recovery rate | 1 week$^{-1}$ |
| $\alpha_H$ | Humans mortality rate due to the disease | zero |
| $\beta_{ij}^M$ | Proportion of mosquitoes from neighborhood $i$ that bite humans from neighborhood $j$ | variable |
| $\mu_M$ | Mosquitoes mortality rate | 0.33 week$^{-1}$ |
| $\tau$ | Extrinsic incubation period of dengue virus | 1 week |
| $\varphi$ | Probability of survival through the extrinsic incubation period* | 0.74 |
| $\Lambda_i^M$ | Growing rate of Mosquitoes | Variable (usually zero) |

* $\varphi$ is usually taken to be $\exp(-\mu_M \tau)$

The equations that describe the system are:

$$\frac{dH_{Si}}{dt} = -a_1 b H_{Si} \sum_j \beta_{ij}^H \frac{M_{Ij}}{N_{Hj}} - \mu_H H_{Si} + \Lambda_i^H$$

$$\frac{dH_{Ii}}{dt} = a_1 b H_{Si} \sum_j \beta_{ij}^H \frac{M_{Ij}}{N_{Hj}} - (\mu_H + \gamma_H + \alpha_H) H_{Ii}$$

$$\frac{dH_{Ri}}{dt} = \gamma_H H_{Ii} - \mu_H H_{Ri}$$

$$\frac{dM_{Si}}{dt} = -a_2 c \frac{M_{Si}}{N_{Hi}} \sum_j \beta_{ij}^M H_{Ij} - \mu_M M_{Si} + \Lambda_i^M \qquad (1)$$

$$\frac{dM_{Li}}{dt} = a_2 c \frac{M_{Si}}{N_{Hi}} \sum_j \beta_{ij}^M H_{Ij} - a_2 c \varphi \frac{M_{Si}(t-\tau)}{N_{Hi}(t-\tau)} \sum_j \beta_{ij}^M H_{Ij}(t-\tau) - \mu_M M_{Li}$$

$$\frac{dM_{Ii}}{dt} = a_2 c \varphi \frac{M_{Si}(t-\tau)}{N_{Hi}(t-\tau)} \sum_j \beta_{ij}^M H_{Ij}(t-\tau) - \mu_M M_{Ii}$$

where $i, j = 1 \ldots N$ is the number of neighborhoods, $\varphi$ is the proportion of latent mosquitoes that survived the incubation period $\tau$, and

$$\beta_{ii}^H = 1 - \sum_{j \neq i} \beta_{ij}^H \text{ and } \beta_{ii}^M = 1 - \sum_{j \neq i} \beta_{ij}^M, \text{ so that } \sum_j \beta_{ij}^H = \sum_j \beta_{ij}^M = 1.$$

As an example to clarify the above equations, let us consider two neighborhoods and two human and insect populations. Then, we have:

$$\frac{dH_{S1}}{dt} = -a_1 b H_{S1} \left[ (1-\beta_{12}^H) \frac{M_{I1}}{N_{H1}} + \beta_{12}^H \frac{M_{I2}}{N_{H2}} \right] - \mu_H H_{S1} + \Lambda_1^H$$

$$\frac{dH_{S2}}{dt} = -a_1 b H_{S2} \left[ \beta_{21}^H \frac{M_{I1}}{N_{H1}} + (1-\beta_{21}^H) \frac{M_{I2}}{N_{H2}} \right] - \mu_H H_{S2} + \Lambda_2^H$$

$$\frac{dH_{I1}}{dt} = a_1 b H_{S1} \left[ (1-\beta_{12}^H) \frac{M_{I1}}{N_{H1}} + \beta_{12}^H \frac{M_{I2}}{N_{H2}} \right] - (\mu_H + \gamma_H + \alpha_H) H_{I1}$$

$$\frac{dH_{I2}}{dt} = a_1 b H_{S2} \left[ \beta_{21}^H \frac{M_{I1}}{N_{H1}} + (1-\beta_{21}^H) \frac{M_{I2}}{N_{H2}} \right] - (\mu_H + \gamma_H + \alpha_H) H_{I2}$$

$$\frac{dH_{R1}}{dt} = \gamma_H H_{I1} - \mu_H H_{R1}$$

$$\frac{dH_{R2}}{dt} = \gamma_H H_{I2} - \mu_H H_{R2}$$

and

$$\frac{dM_{S1}}{dt} = -a_2 c \frac{M_{S1}}{N_{H1}}\left[(1-\beta_{12}^M)H_{I1}+\beta_{12}^M H_{I2}\right]-\mu_M M_{S1}+\Lambda_1^M$$

$$\frac{dM_{S2}}{dt} = -a_2 c \frac{M_{S2}}{N_{H2}}\left[\beta_{21}^M H_{I1}+(1-\beta_{21}^M)H_{I2}\right]-\mu_M M_{S2}+\Lambda_2^M$$

$$\frac{dM_{L1}}{dt} = a_2 c \frac{M_{S1}}{N_{H1}}\left[(1-\beta_{12}^M)H_{I1}+\beta_{12}^M H_{I2}\right]-$$

$$a_2 c \varphi \frac{M_{S1}(t-\tau)}{N_{H1}(t-\tau)}\left[(1-\beta_{12}^M)H_{I1}(t-\tau)+\beta_{12}^M H_{I2}(t-\tau)\right]-\mu_M M_{L1} \quad (2)$$

$$\frac{dM_{L2}}{dt} = a_2 c \frac{M_{S2}}{N_{H2}}\left[\beta_{21}^M H_{I1}+(1-\beta_{21}^M)H_{I2}\right]-$$

$$a_2 c \varphi \frac{M_{S2}(t-\tau)}{N_{H2}(t-\tau)}\left[\beta_{21}^M H_{I1}(t-\tau)+(1-\beta_{21}^M)H_{I2}(t-\tau)\right]-\mu_M M_{L2}$$

$$\frac{dM_{I1}}{dt} = a_2 c \varphi \frac{M_{S1}(t-\tau)}{N_{H1}(t-\tau)}\left[(1-\beta_{12}^M)H_{I1}(t-\tau)+\beta_{12}^M H_{I2}(t-\tau)\right]-\mu_M M_{I1}$$

$$\frac{dM_{I2}}{dt} = a_2 c \varphi \frac{M_{S2}(t-\tau)}{N_{H2}(t-\tau)}\left[\beta_{21}^M H_{I1}(t-\tau)+(1-\beta_{21}^M)H_{I2}(t-\tau)\right]-\mu_M M_{I2}$$

From the above equations, it becomes clear that only individuals of the human populations travel from one neighborhood to another. For example, the proportion of susceptible and infected humans from neighborhood 1 that visit neighborhood 2 is denoted $\beta_{12}^H$. In addition we assume that $\beta_{ij}^H \ (i \neq j)$ and $\beta_{ij}^M \ (i \neq j)$ are small.

Seasonal fluctuations in the mosquitoes number is very common and may influence the pattern and the duration of outbreaks. To introduce these fluctuations we replace the mosquitoes equations by:

$$\frac{dM_{S1}}{dt} = \left[1 + M_2^1\left(\frac{2\pi f}{\mu_M}\right)\cos(2\pi f t)\right]\mu_M(M_{S1} + M_{L1} + M_{I1}) -$$

$$a_2 c \frac{M_{S1}}{N_{H1}}\left[(1-\beta_{12}^M)H_{I1} + \beta_{12}^M H_{I2}\right] - \mu_M M_{S1}, \qquad M_2^1\left(\frac{2\pi f}{\mu_M}\right) < 1$$

$$\frac{dM_{S2}}{dt} = \left[1 + M_2^2\left(\frac{2\pi f}{\mu_M}\right)\cos(2\pi f t)\right]\mu_M(M_{S1} + M_{L1} + M_{I1}) -$$

$$a_2 c \frac{M_{S2}}{N_{H2}}\left[\beta_{21}^M H_{I1} + (1-\beta_{21}^M)H_{I2}\right] - \mu_M M_{S2}, \qquad M_2^2\left(\frac{2\pi f}{\mu_M}\right) < 1$$

$$\frac{dM_{L1}}{dt} = a_2 c \frac{M_{S1}}{N_{H1}}\left[(1-\beta_{12}^M)H_{I1} + \beta_{12}^M H_{I2}\right] -$$

$$a_2 c \varphi \frac{M_{S1}(t-\tau)}{N_{H1}(t-\tau)}\left[(1-\beta_{12}^M)H_{I1}(t-\tau) + \beta_{12}^M H_{I2}(t-\tau)\right] - \mu_M M_{L1} \qquad (3)$$

$$\frac{dM_{L2}}{dt} = a_2 c \frac{M_{S2}}{N_{H2}}\left[\beta_{21}^M H_{I1} + (1-\beta_{21}^M)H_{I2}\right]$$

$$- a_2 c \varphi \frac{M_{S2}(t-\tau)}{N_{H2}(t-\tau)}\left[\beta_{21}^M H_{I1}(t-\tau) + (1-\beta_{21}^M)H_{I2}(t-\tau)\right] - \mu_M M_{L2}$$

$$\frac{dM_{I1}}{dt} = a_2 c \varphi \frac{M_{S1}(t-\tau)}{N_{H1}(t-\tau)}\left[(1-\beta_{12}^M)H_{I1}(t-\tau) + \beta_{12}^M H_{I2}(t-\tau)\right] - \mu_M M_{I1}$$

$$\frac{dM_{I2}}{dt} = a_2 c \varphi \frac{M_{S2}(t-\tau)}{N_{H2}(t-\tau)}\left[\beta_{21}^M H_{I1}(t-\tau) + (1-\beta_{21}^M)H_{I2}(t-\tau)\right] - \mu_M M_{I2}$$

$$M_{L1}(0) = M_{L2}(0) = 1$$
$$M_{I1}(0) = M_{I2}(0) = 0$$

In system (3) $M_2^1$ and $M_2^2$ define the amplitude of the oscillations in the density of mosquitoes due to seasonality (this is a schematic model). In the absence of infection, the equation for the mosquitoes populations can be solved exactly. The equation for the density of mosquitoes and its solutions are:

$$\frac{dM_{Si}}{dt} = \left[M_2^i\left(\frac{2\pi f}{\mu_M}\right)\cos(2\pi f t)\right]\mu_M(M_{Si}), \quad i = 1,2 \qquad (3a)$$

and

$$M_{Si} = M_{Si}(0)\exp\left[M_2^i \sin(2\pi f t)\right] \qquad (3b)$$

Note that, in system (3), the total number of mosquitoes is not affected by the infection. This is a very good approximation for dengue because only a very small amount of mosquitoes is infected and their life expectation is not affected by the disease. However, the total number of infections varies with the climatic factor as compared with the case in which the number of mosquitoes is constant. It also varies with the instant of time the infection is "introduced" in the population.

If the disease is introduced, say, in the district 1 it will eventually spread to other districts. This however can be a very slow process and since that data reported is the number of cases in a given city per month, duration of the epidemics can be very long because the disease is "travelling" through the city carried by Humans.

**Remark 1:** Choosing $\Lambda_i^H$ and $\Lambda_i^M$ in such a way that the Human and the Mosquito populations are kept constant, then dividing the equations of system (1) by $N_{Hi}$ and $N_{Mi}$ we get equations for the proportions. Of course if there is seasonality, this cannot be done for the mosquito population.

**Remark 2:** The above equations refers to a single dengue strain circulating in a given population. In some places, however, more than one serotype can circulate simultaneously in the same community. If we assume that the circulation of one virus does not interfere in the other then the two virus can be treated separately.

**Remark 3:** In this paper we assumed two different biting rates (denoted $a_1$ and $a_2$) for the infected and non-infected mosquitoes, respectively. We did this for the sake of generality. For the case of dengue it is accepted that they are both equal but this may not be true for other vector-borne infections (for instance, the case of plague [5])

## 2. Analysis of the model

The system of equations given above does not have an analytical solution even when we have just one neighborhood and no seasonal fluctuations. Even if an analytical solution would exist it would be so complicated that it would be useless. We, therefore, list a number of numerical results about the system and show how they can be used.

### 2.1. The simple Ross-Macdonald model

Let us consider that we have just one human and one mosquito population. As pointed out in Amaku et al. [6], this simple system is solved by assuming an initial population of infected (either humans or mosquitoes) individuals uniformly distributed over the area we are studying. The result is a huge outbreak, as shown in figure(1). This can be explained as follows. If the basic reproduction number [7-9]:

$$R_0 = \frac{a_1 a_2 bc\varphi}{\mu_M (\mu_H + \gamma_H + \alpha_H)} \frac{N_M(0)}{N_H(0)} \tag{4}$$

of this model is greater than 1, the outbreak increases up to a certain point where the number of susceptible diminishes to a certain value. The infection then disappears for a number of years. This is not what is observed in real outbreaks of dengue, as discussed in section 1. This will be elaborated latter in this paper.

To determine when the number of new cases begins to diminish, we use the following approximate threshold [10]:

$$Th(t) = \frac{\varphi a_1 a_2 bc}{\mu_M (\mu_H + \gamma_H + \alpha_H)} \frac{M_S(t)}{N_H(t)} \frac{H_S(t)}{N_H(t)} = R_0 \frac{M_S(t)}{N_H(t)} \frac{H_S(t)}{N_H(t)} \frac{N_H(0)}{N_M(0)} \quad (5)$$

When $R(t)$ crosses the unit from above (below), the incidence and prevalence reach a maximum (minimum). We consider only the first outbreak. that is, the only one that has physical meaning.

It is important to note three things:

1) If nothing changes in the neighborhood we are studying, then when the threshold, $Th(t)$, is reached, the outbreak is interrupted and the infection disappears rapidly. The number of residual susceptibles, $H'_S$, results in a new $R_0$ that we call $R_{0b} = R_0 H'_S/N_H$ which is less than 1 in the following year. Therefore, no outbreak is expected with likelihood that is numerically equal to this new $R_{0b}$ (see discussion below). On the other hand, if something changes during the outbreak (heavy rains, public health interventions, drought, etc), then the threshold may be reached, for example by variation of the factor $\frac{M_S(t)}{N_H(t)}$ in equation (5). In this case, the remaining number of susceptibles is large enough and we expect another outbreak in the following year since the population of mosquitoes increases very rapidly (see discussion below).

2) The number of cases predicted by the model, if nothing changes, is much larger than the number of cases actually observed and reported in all dengue outbreaks we know. At first sight this is strange because the parameters used in the model are obtained empirically. An explanation for this will be given later.

3) The total number of cases in the first outbreak predicted by this simple model increases with the two components of $R_0$ [11]:

$$T_{H \to M} = \frac{N_V a_2 c}{N_H (\gamma_H + \mu_H + \alpha_H)} \quad (5),$$

$$T_{M \to H} = \frac{a_1 b \varphi}{\mu_M} \quad (6),$$

with

$$R_0 = T_{H \to M} \times T_{M \to H} \tag{7}$$

as shown in Tables 2a, 2b and 2c.

Table 2a: Percentage of infected individuals in the first outbreak.

| $T_{H \to M}/T_{M \to H}$ | 0.71 | 0.80 | 0.89 | 0.98 | 1.06 | 1.15 | 1.24 | 1.33 | 1.42 |
|---|---|---|---|---|---|---|---|---|---|
| 1.30 | 0.00 | 2.24 | 25.20 | 38.51 | 48.77 | 56.86 | 63.35 | 68.63 | 72.98 |
| 1.46 | 2.24 | 27.02 | 41.21 | 51.86 | 60.07 | 66.54 | 71.73 | 75.94 | 79.39 |
| 1.62 | 25.11 | 41.09 | 52.71 | 61.44 | 68.16 | 73.46 | 77.69 | 81.11 | 83.92 |
| 1.79 | 38.21 | 51.54 | 61.23 | 68.52 | 74.13 | 78.54 | 82.05 | 84.90 | 87.22 |
| 1.95 | 48.19 | 59.48 | 67.71 | 73.89 | 78.65 | 82.37 | 85.34 | 87.72 | 89.67 |
| 2.11 | 55.95 | 65.66 | 72.74 | 78.05 | 82.13 | 85.32 | 87.85 | 89.89 | 91.54 |
| 2.27 | 62.10 | 70.56 | 76.71 | 81.33 | 84.87 | 87.63 | 89.82 | 91.57 | 92.98 |
| 2.44 | 67.06 | 74.49 | 79.90 | 83.95 | 87.05 | 89.47 | 91.37 | 92.89 | 94.12 |
| 2.60 | 71.11 | 77.70 | 82.49 | 86.08 | 88.82 | 90.95 | 92.62 | 93.95 | 95.03 |

Table 2b: Percentage of latent mosquitoes in the first outbreak.

| $T_{H \to M}/T_{M \to H}$ | 0.71 | 0.80 | 0.89 | 0.98 | 1.06 | 1.15 | 1.24 | 1.33 | 1.42 |
|---|---|---|---|---|---|---|---|---|---|
| 1.30 | 0.00 | 0.14 | 1.56 | 2.37 | 2.99 | 3.46 | 3.84 | 4.14 | 4.38 |
| 1.46 | 0.15 | 1.87 | 2.85 | 3.56 | 4.10 | 4.51 | 4.83 | 5.08 | 5.28 |
| 1.62 | 1.94 | 3.15 | 4.01 | 4.64 | 5.10 | 5.46 | 5.73 | 5.94 | 6.10 |
| 1.79 | 3.22 | 4.31 | 5.07 | 5.62 | 6.03 | 6.33 | 6.56 | 6.73 | 6.86 |
| 1.95 | 4.40 | 5.37 | 6.05 | 6.53 | 6.88 | 7.13 | 7.32 | 7.46 | 7.57 |
| 2.11 | 5.49 | 6.36 | 6.95 | 7.37 | 7.67 | 7.88 | 8.03 | 8.14 | 8.22 |
| 2.27 | 6.49 | 7.27 | 7.79 | 8.15 | 8.40 | 8.58 | 8.70 | 8.78 | 8.84 |
| 2.44 | 7.43 | 8.12 | 8.58 | 8.88 | 9.09 | 9.23 | 9.32 | 9.38 | 9.41 |
| 2.60 | 8.30 | 8.91 | 9.31 | 9.57 | 9.74 | 9.85 | 9.91 | 9.94 | 9.95 |

Table 2c: Percentage of infected mosquitoes in the first outbreak.

| $T_{H \to M}/T_{M \to H}$ | 0.71 | 0.80 | 0.89 | 0.98 | 1.06 | 1.15 | 1.24 | 1.33 | 1.42 |
|---|---|---|---|---|---|---|---|---|---|
| 1.30 | 0.00 | 0.10 | 1.12 | 1.70 | 2.15 | 2.49 | 2.76 | 2.97 | 3.15 |
| 1.46 | 0.11 | 1.35 | 2.05 | 2.56 | 2.95 | 3.24 | 3.47 | 3.65 | 3.80 |
| 1.62 | 1.39 | 2.27 | 2.88 | 3.33 | 3.67 | 3.92 | 4.12 | 4.27 | 4.39 |
| 1.79 | 2.32 | 3.10 | 3.65 | 4.04 | 4.33 | 4.55 | 4.72 | 4.84 | 4.93 |
| **1.95** | 3.17 | 3.86 | 4.35 | 4.70 | 4.95 | 5.13 | 5.27 | 5.37 | 5.44 |
| **2.11** | 3.95 | 4.57 | 5.00 | 5.30 | 5.51 | 5.67 | 5.78 | 5.86 | 5.91 |
| **2.27** | 4.67 | 5.23 | 5.60 | 5.86 | 6.04 | 6.17 | 6.26 | 6.32 | 6.35 |
| **2.44** | 5.34 | 5.84 | 6.17 | 6.39 | 6.54 | 6.64 | 6.70 | 6.75 | 6.77 |
| **2.60** | 5.97 | 6.41 | 6.70 | 6.88 | 7.00 | 7.08 | 7.13 | 7.15 | 7.16 |

The epidemics for large $R_0$ eventually saturates the population, that is the proportion of infected people approaches one, as shown in Figure 2.

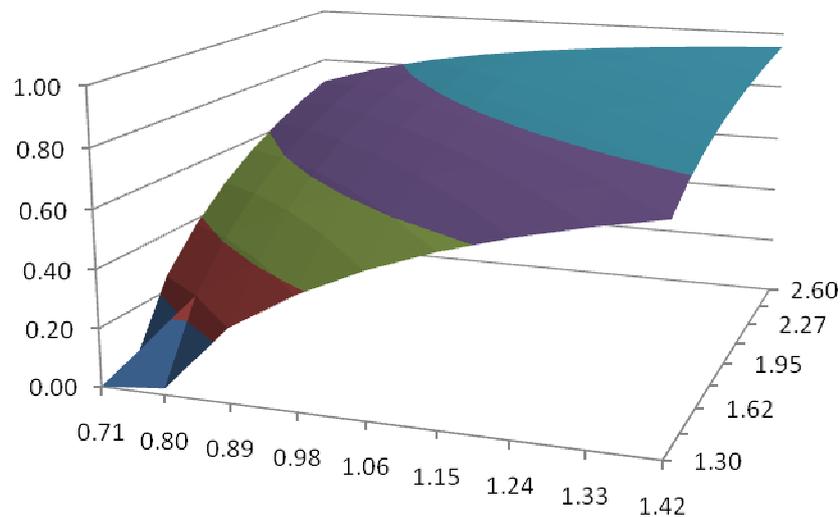

Figure 2: Proportion of infected individuals in the first outbreak as a function of $T_{H \to M}$ and $T_{M \to H}$.

This result implies that for large values of $R_0$, which is the product of $T_{H \to M}$ and $T_{M \to H}$, it is very unlikely that the infection will return in the subsequent year after the outbreak. We will return to this point later on the paper.

**2.2. Simulating some observed patterns of dengue recurrence**

The epidemic patterns (time-distribution of the yearly incidence of the infection) in many cities of Brazil are extremely complicated. However, the model given by the system of equations (1) can be used to fit any observed pattern. In the results section we show two examples of such fitting for two different cities with completely different patterns: Natal and Recife (both at North-East of Brazil)

Let us explain here how this can be done. As explained above the classical Ross-Macdonald model, that is, just one population of humans and one population of mosquitoes with $R_0 > 1$, produces a huge outbreak followed by smaller blips of infection that are widely separated in time (typically decades) and, therefore, non-physical. This huge outbreaks occurs because, as pointed out in Amaku et al. [6], this equation is solved by assuming an initial population of infected individuals uniformly distributed over the region being studied. Therefore, the disease occurs simultaneously everywhere over the area and produces a single very large outbreak, as shown in figure 3. After this peak, the infection dies out, that is, the number of cases drops down to zero, remaining close to zero for several years. This happens when the threshold [10] is reached and this occurs even when $R_0$ is only slightly greater than 1.

As mentioned above, however, other factors may cause $Th(t)$ to fall below 1. For instance, factors affecting mosquitoes survival, like heavy rains, droughts, sanitary interventions, cold weather, may result in $Th(t) < 1$. As a consequence, it is possible to have a residual $H'_S(t)/N_H(t)$ after the outbreak large enough that should guarantee an outbreak in the following year but, due to a smaller mosquitoes densities, the outbreak may not occur. Summarizing, if external factors do not interfere with the intensity of transmission, the outbreak is interrupted by the lack of enough susceptibles to maintain it and we will have a residual $H'_S(t)/N_H(t)$ low enough that an outbreak in the following year is rendered unlikely. If, however, the outbreak is interrupted by external factors before $H'_S(t)/N_H(t)$ is low enough, then the new $R_{0b}$ will be greater than 1, and then an outbreak in the following year will possibly occur. The size of the new outbreak will be dependent on the number of human susceptibles available to be infected and may be larger or smaller than the first outbreak.

If we subdivide the populations into several sub-populations and introduce the disease in just one of them, then we have as many peaks as the number of sub-populations. In this case, the total population is known as 'meta-population' in the literature [11]. Figure 4 illustrates the pattern obtained with three identical sub-populations.

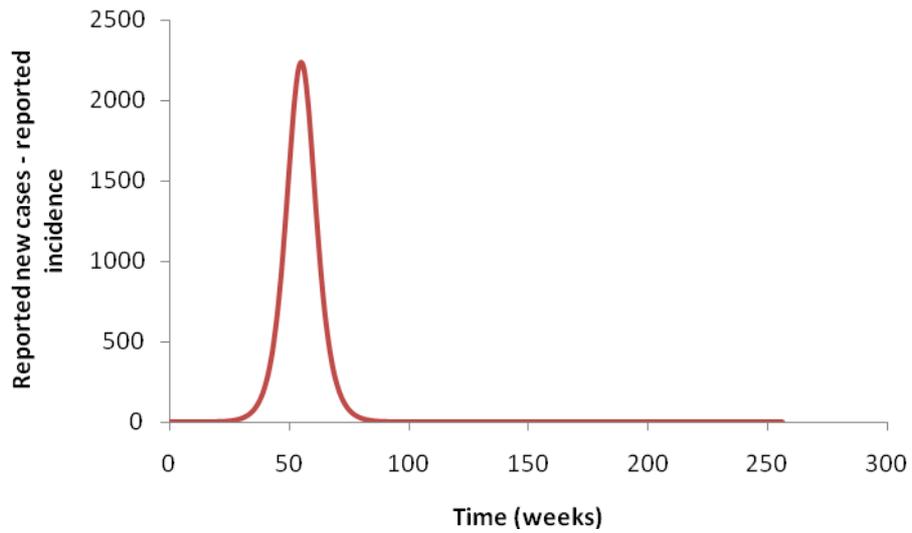

Figure 3: Simulation of an outbreak when the initial condition is spread all over the area: we get a single large peak.

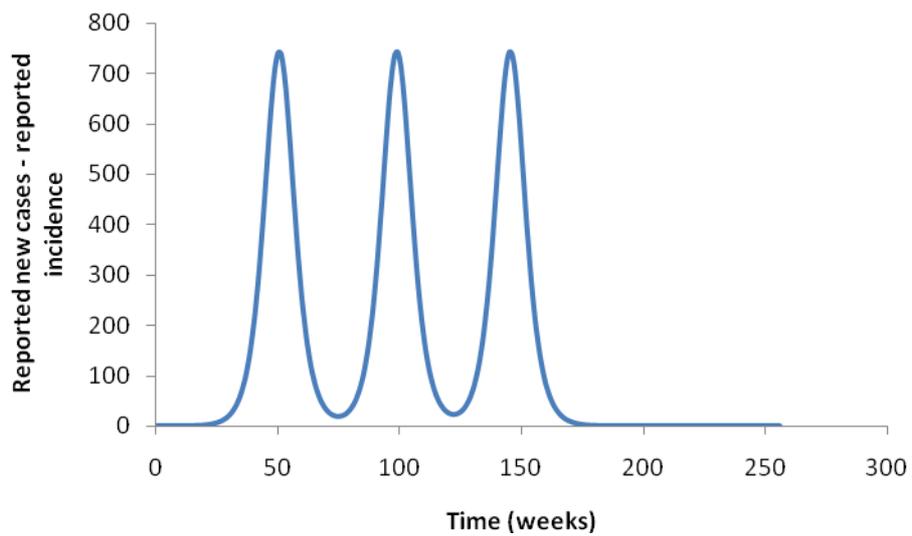

Figure 4: Simulated pattern obtained with 3 identical sub-populations geographically separated. The infection was introduced in the first sub-population and propagated to the others.

Introducing seasonality completely changes the pattern of figure 4, as illustrated in figure 5.

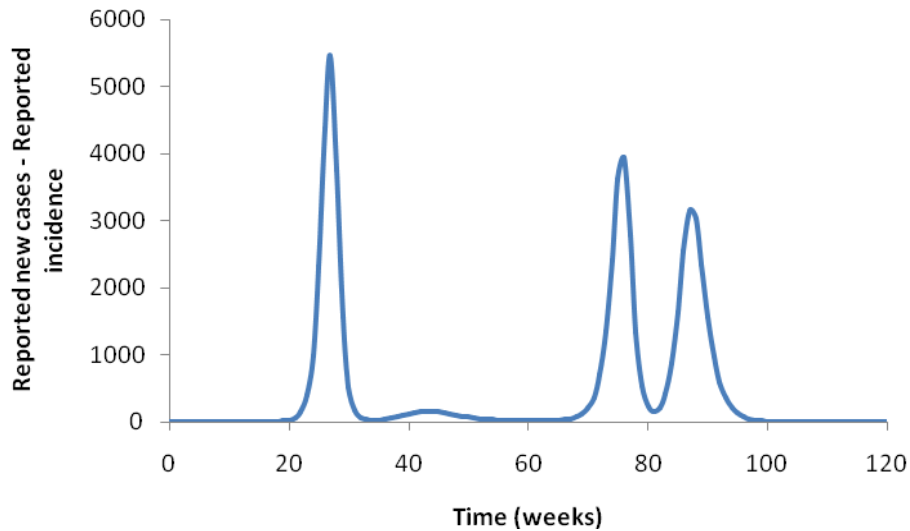

Figure 5: The effect of seasonality on the pattern shown in Figure 4.

### 2.3. An important observation: the invariance of the total number of cases after the first outbreak

It is important to note that the total number of cases is independent of the pattern of outbreaks (time-distribution of incidence of cases) if there is no seasonality. It rather depends on the values of the parameter that determine the intensity of transmission at the beginning of the epidemic. In fact, the number of cases in the single outbreak produced by the simple Ross-Macdonald model is almost the same as the total number of cases resulting from summing up the 3 peaks of the case illustrated by figure 4. The small difference occurs because the single peak of the Ross-Macdonald model has a small duration in time, whereas in the other cases, the outbreaks last for a few years and, therefore, demographic effects(people are born and die) take place in between. Obviously, when the mortalities and birth rates are negligible, the total number of cases in the first and only outbreak is independent of the pattern of dengue recurrence.

When the basic reproduction number is too low (only slightly above unit), however, although the threshold given by equation (5) is reached, the number of remaining susceptible after the outbreak may be large (see figure 2).

### 2.4. The effect of seasonality

When seasonality is considered, the total number of mosquitoes is not constant anymore and can vary considerably along the year. In figure 5 we introduced seasonality to modify the patterns shown in figure 4. The total number of cases after 3 years of simulation is quite higher than the one in the case where no seasonality is considered. The total number of cases depends on the time of the year the infection is 'introduced'. In figure 6 we show the variation in the number of cases as a function of the moment the infection is introduced for four different mosquitoes-to-humans ratios. In the figure, the dotted lines represent the averages around which the number of cases oscillate.

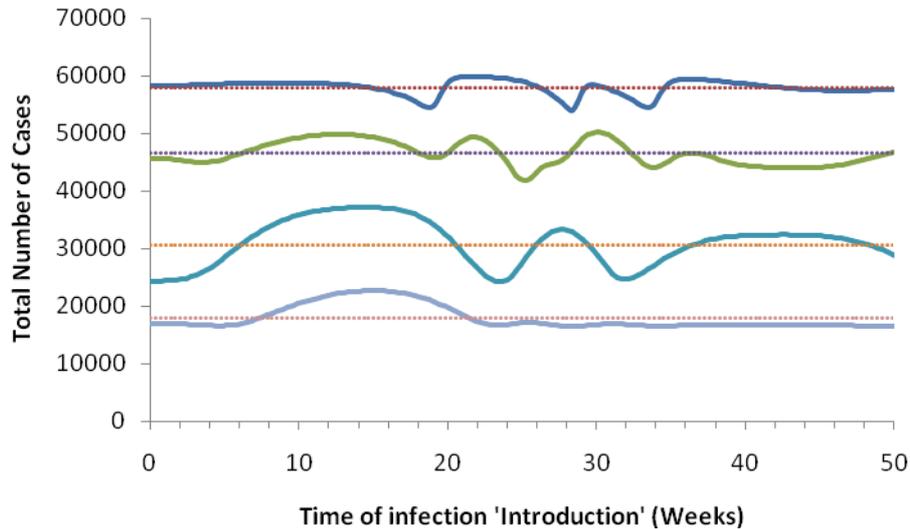

Figure 6: The effect of seasonality in the total number of cases for 4 mosquitoes-to-humans ratios. These ratios varied from the ranges [0.225-3.6] for the bottom lines to [0.5-8.0] for the uppermost line. The continuous lines represent the total number of cases when seasonality is considered. They oscillate around the averages (dotted lines) depending on the time of the infection is 'introduced'.

Although we do not know for sure whether there is any preferential moment for the infection to be introduced in a non-affected area by an infected individual, one should expect that there is a higher probability of introduction in periods of higher infection incidence. Note that the number of cases depends on this moment and, therefore, just this suffice to influence the number of cases if the number of human susceptibles is high enough (above the threshold).

## 2.5. The Asymptomatic Cases: Dark Matter

Using the accepted values for the parameters that enter the Ross-Macdonald model for dengue, we always get a number of infections that is much greater than what has been reported in any endemic area [4]. In this paper, we are going to assume that a proportion of those cases is not identified and/or notified as dengue cases (asymptomatic/undiagnosed acute febrile cases). This proportion varies in the literature from 1:3 to 1:13 [4], [13]. We call this unreported/unidentified cases "dark matter", following the cosmologists jargon. In the examples that follow, we calculate this proportion in some locations. Unfortunately, we have only one place in Brazil where both seroprevalence data and dengue reported cases exist [14].

## 2.5. More than one strain of dengue virus or different viruses (Chikungunya, Zika, Yellow Fever, etc)

If there is super-infection by other dengue strains of viruses in either humans or mosquitoes, there are two distinct possibilities: 1) no competition between strains or viruses. In this case, each outbreak can be treated separately. For example, if Zika virus do not compete with dengue or other virus, then we can predict that the outbreak of Zika will be very similar to the outbreak of dengue in the same region; 2) competition between strains or viruses. In this case, an entirely new calculation has to be performed. A possible mechanism for competition can be found in [15], [16] and will be used in a future paper.

## 3. Results: Understanding the dengue oscillator

The main qualitative results of this paper are the following.

### 3.1 An outbreak in a limited district

If the number of mosquitoes in an area of a certain district is such that the $R_0$ of this area is greater than 1 the disease may invade and an outbreak occurs. In this case, the incidence of the disease will increase until the proportion of susceptible in the host population decreases to a certain level approximately given by [10]:

$$\frac{H_{Si}}{N_{Hi}} = \frac{1}{R_0}\left(\frac{N_{Mi}}{M_{Si}}\right). \tag{8}$$

Then the outbreak will decrease and disappear from this district for a number of years. Immediately after the outbreak, the proportion of remaining susceptible can be calculated as 1 minus the values shown in table 2a.

On the other hand, if for some reason after the outbreak the new $R_{0b}$ is increased at least by a factor $N_{Hi}/H_{Si}$, then another outbreak may occur. In this case, depending on the remaining proportion of susceptible and the new $R_{0b}$ the number of affected by the new outbreak may be larger or smaller than the first one and the number of resistant individuals will approach saturation.

The important point is that, when $R_0$ of this area is sufficiently above 1, the total number of cases among humans turns out to be independent of variations in the transmission components $T_{H \to M}$ and $T_{M \to H}$, provided their values are not significantly altered. For example, if the mosquito density with respect to humans is lower in one district than in the other, then the outbreak will take more time to disappear but the number of cases will be smaller, see table 2a. Note that any outbreak occurring afterwards will depend on the increase of the new $R_{0b}$ after the outbreak.

Using these results we can attempt to make some predictions at a certain point of the epidemics of what will happen in the near future. To exemplify: suppose we have a given district with intensity of transmission such that $R_{0b}$ for this district is greater than one. Then one can calculate what will be the number of cases reported in this district at the end of the outbreak. If at some point of the epidemics the total number of cases is above a certain threshold, we can say that the epidemics is over, or, with great probability, if it will continue in time and how many more case it will have.

## 3.2. Examples of Calculation

The calculation reported below follow these steps:

a) looking at the data, we tentatively identify epidemics that presumably were interrupted by reaching the threshold given by equation (5). These epidemics are identified by long periods of very low dengue activity between two successive outbreaks.

b) The epidemics may consist of outbreaks that last a few years but if they recur for more than five consecutive years the calculation described below breaks down.

c) The calculation assumes a proportion (1- $\eta$) of dark matter and the epidemics is considered over when the number of susceptibles drops below a certain threshold (see Equation 5). Of course $\eta$, the proportion of notified cases, should be confirmed by actual observations. However, the examples below seem to indicate that for a given geographical region the value of $\eta$ do not vary very much from place to place.

*3.2.1. A single outbreak: The case of Recife dengue outbreak 2001-2002*

This outbreak consisted in only one peak of cases followed by several years with only marginal dengue transmission. We can then tentatively assume that the number of susceptible individuals dropped to a level that interrupted the transmission due to herd immunity. In Figure 6, we fitted this outbreak, calculate the total number of cases and found out that only assuming a value of 1:25 for the dark matter we recover the notification data.

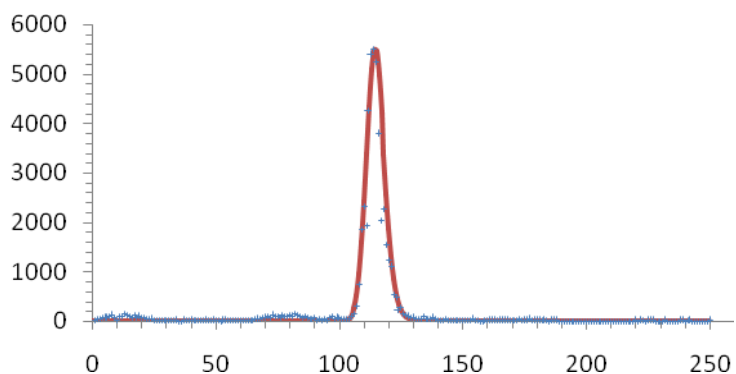

Figure 6: Number of weekly real (crosses) and calculated (continuous line) notified cases for the 2001-2002 outbreak in Recife, Brazil. The total number of notified cases was 48,500. The total number of cases given by the model, considering $\eta = 0.04$ is 1,260,000.

This enormous number of cases represents more than 85% of the total Recife population at the time indicating that the threshold was reached. Incidentally one could predict that this particular strain of dengue virus will take some time to return. This in fact happened. On the other hand, this proportion agrees with the survey carried out by Braga *et al*. [14], and a herd immunity corresponding to $R_0 > 4$.

*3.2.2. Multiple outbreaks: The case of Natal dengue outbreak 2000-2007*

Contrasting with the single Recife outbreak described above, dengue in Natal in the period between 2000 and 2007 presented a recurrent pattern with annual outbreaks of smaller magnitude. This pattern was reproduced by the model using nine geographical regions (neighborhoods) and assuming that the infection propagates through the movement of infected humans hosts among those regions (see equation (1)). The period between weeks 200 and 330 was fitted by reducing the mosquitoes density. Of course, there are other mechanisms that could be used to fit this pattern. The important thing is that the number of asymptomatic cases (dark matter) added to the reported cases is greater than Natal total population. This can be explained by the fact that there are more than one strain of dengue virus circulating and that the population increased 1% per year during the period. We shall return to this point later in the paper.

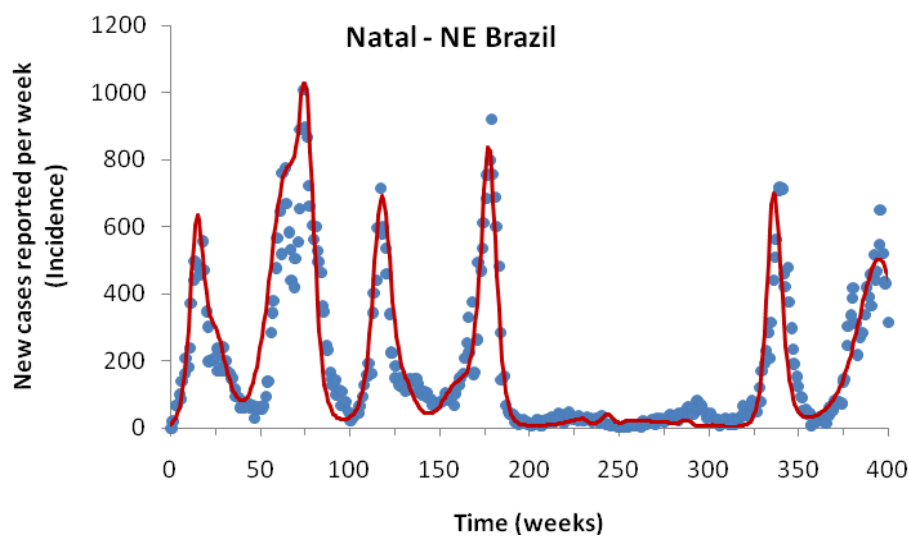

Figure 7. Simulation of the outbreaks in Natal, NE of Brazil.

*3.2.3 Forecasting outbreaks in a small neighborhood*

Consider a small homogeneous neighborhood. In order to predict outbreaks and their intensity, we need basically three quantities: transmission components $T_{H \to M}$ and $T_{M \to H}$ of the neighborhood and the value of $\eta$, which gives the proportion of dark matter (asymptomatic cases). The problem is how to measure or to estimate these three quantities. Suppose we have a neighborhood where an outbreak happened and faded away for few consecutive years. Then, we can proceed in the following way. Compare the fraction of reported cases with the fraction predicted by the model in table 1. If the outbreak occurred a number of years ago it is safe to consider that the value of $\eta$ is big. Assuming several $\eta$'s we can calculate the likelihood of another outbreak and its magnitude.

Unfortunately the value of $\eta$ must be measured and that may be expensive. However, we can assume several values of $\eta$ and calculate the corresponding $R_0$ that explains each $\eta$. The $R_0$ can also be calculated by another methods and then the value of $\eta$ can be deduced.

Even when the outbreak saturates a given neighborhood of a city, new outbreaks of dengue may be reported in the same city. These outbreaks occur in some other neighborhoods due to the movement of people. In the next section, we examine this problem.

*3.2.4. Forecasting outbreaks in large cities*

Example 1: São Paulo (Brazil), 2014-2015.

In this case, due to human movement, there are a number of peaks before the whole outbreak fades out. Of course we have to fit the patterns considering the different neighborhoods but, in order to make predictions it is necessary to know in which neighborhood a particular outbreak occurs. If this is known, then the calculation described in 3.2.3 can be carried out for each region. However, if the city is really large, the outbreaks may be restricted to some boroughs. In this case we have to consider the population of the neighborhoods not affected to calculate the total number of cases that may occur in the future. Unfortunately, without further information, it is impossible to predict the order of the outbreaks in time and space. We are going to exemplify this situation with an extreme case, namely the 2013, 2014 and 2015 dengue outbreaks in the megalopolis São Paulo (Brazil), where we can see that the outbreaks 'travel' from place to place in subsequent years (figure 8).

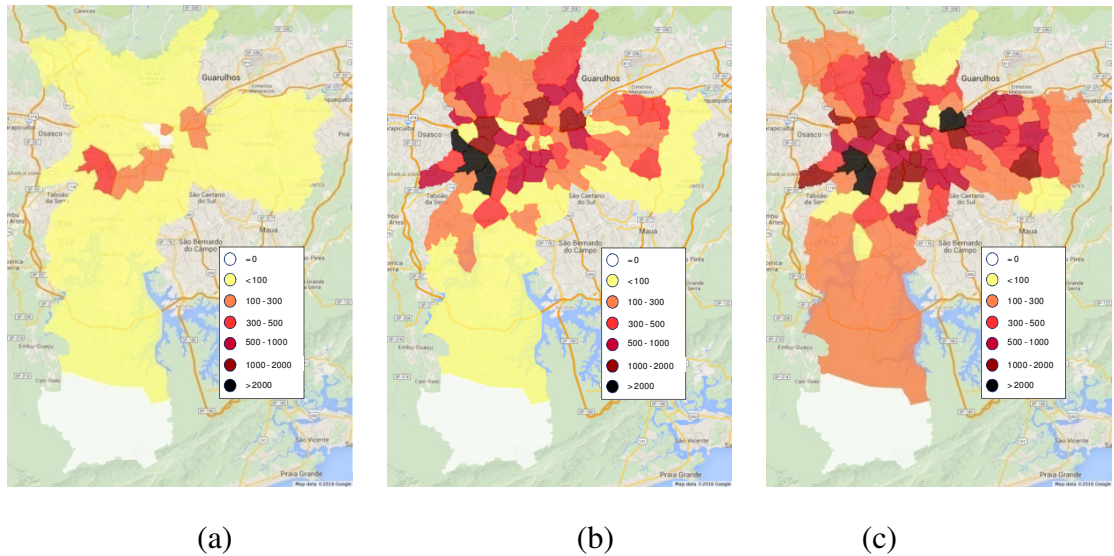

(a)          (b)          (c)

Figure 8: Figures a, b and c show the evolution of the dengue outbreak in São Paulo. Figure a is the situation at the end of 2013, b, 2014 and c, 2015. The colors represent the number of cases per 100,000 inhabitants.

Example 2: Rio de Janeiro (Brazil) 2000, 2008 and 2011-2013.

In this example we illustrate several features of the model.

Rio de Janeiro had three outbreaks in 2000, 2008 and 2011-2013. The two epidemics in 2000 and 2008 consisted essentially of two large peaks. We assume, with some data, that the 2000 outbreak was essentially serotype 3 and the 2008 was mainly serotype 2. The three outbreaks in 2011-2013 are assumed to be serotype 4 and 1 simultaneously.

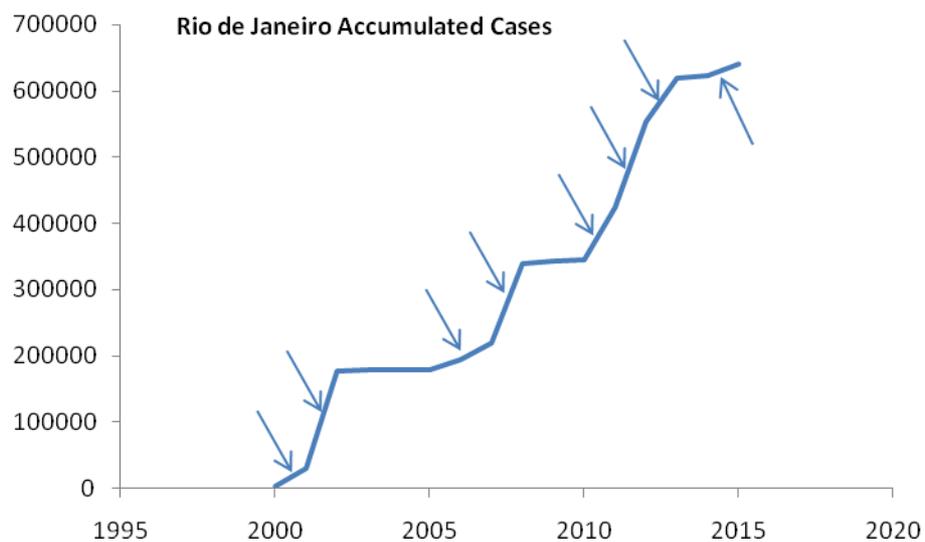

Figure 9: Accumulated reported cases in Rio de Janeiro (Brazil), 2000-2015

The sum of reported cases can be seen in Figure 9. It shows

1. that the outbreaks show different slopes in several parts before reaching a plateau. This can be interpreted as the epidemics moving through the city.
2. The outbreaks of 2000 and 2008 reached approximately the same number of cases. From this we can deduce a value for η.
3. The outbreaks of 2011-2013 did not reach the whole city. In figure 10a and 10b we show the accumulated number of cases for two different districts of Rio, namely, Ilha do Governador and Centro.

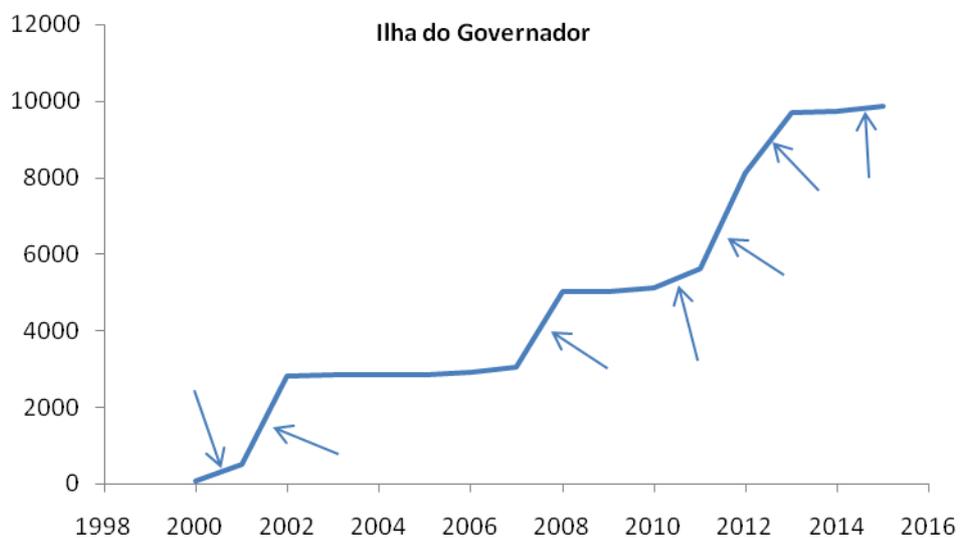

Figure 10a. Accumulated number of dengue cases in the Ilha do Governador borough. Arrows point to epidemic spreading to other regions of the borough. The outbreaks of 2011-2013 are approximately double of the previous outbreaks. This indicates that there were two strains circulating.

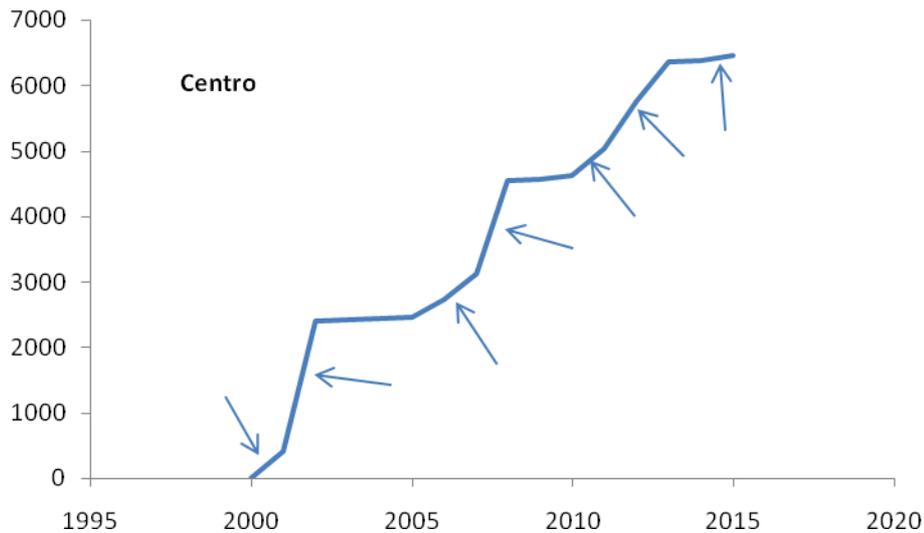

Figure 10b. Accumulated number of dengue cases in the central region of Rio de Janeiro. Arrows point to epidemic spreading to other regions of the borough. Note that in this case the outbreaks of 2011-2013 are approximately of the same magnitude of the previous outbreaks. This indicates that there was only one strain circulating in this region.

**4. Forecasting**

Forecasting outbreaks in large cities, as we have seen above, may be very difficult. Let us recapitulate the reasons for this:

4.2.1. When an epidemic enters in a large city it will enter in a small neighborhood and then propagates to the rest of the city. This movement can lead to peaks in the same city for three to four years before the epidemic fades out, that is, when condition given by equation (5) crosses one everywhere in the city;

4.2.2. Large cities are rarely homogenous and therefore the interruption of epidemics may occur with different proportions of immune people in different parts of the city;

4.2.3. Even in a given neighborhood, as explained before, the epidemics can be interrupted by climatic factors. If this happens there will be a number of peaks until the threshold given by equation (5) crosses one downwards from above and then the infection disappears.

Therefore, to predict epidemics in large cities one should proceed as follows:

1) choose a well described past outbreak and divide the city into homogeneous regions with respect to the number of cases observed, or better, with respect to the proportion of remaining susceptibles in the end of the outbreak;

2) for each region, calculate the remaining number of susceptible. The difficulty here is to know the exact number of unapparent infections. If this is known, then the total number of cases can be adjusted;

3) if this is not known then we have to choose neighborhoods free from infection for a few years, and so:

3.1) assume values of $\eta$ and calculate $R_0$ for these regions. Of course the value of $\eta$ cannot be too big because then we end up with more people infected than the total population.

3.2) given the population size of this region one can calculate the maximum value of $R_0$ for this particular region.

3.3) for each $R_0$ one can estimate the likelihood of an outbreak in the following year and its maximum size.

## 5 Summary and Conclusions

The number of dengue cases in a city (or in a borough of a city) is reported weekly in the whole country for years. The reported data forms a time series and show annual outbreaks (sometimes more than one outbreak in a single year) of very irregular nature. As mentioned before, we call these series the dengue oscillator.

The purpose of this paper is to explain these outbreaks from a semi-quantitative point of view. To do this we used the classical Ross-Macdonald model. This model, as formulated usually, predicts a huge outbreak followed by smaller outbreaks that occur decades later and that are clearly unphysical. The reason why the model naively applied is unphysical is the initial conditions assume that the infections enters the city or borough uniformly, as pointed out by Amaku et al. [6]. The model also predicts a much large number of cases than is actually observed, although it uses measured parameters. In this paper we show that by dividing the population in small areas and introducing the infection in just one (or in just a few), and introducing seasonality, we are able to reproduce any pattern observed. The true vector of dengue is the human host who carries the infection around the city. The huge number of cases is explained by asymptomatic cases that ranges from a factor 4 to 13 with respect to symptomatic cases.

Understanding the mechanism of dengue transmission is the first step to forecast future outbreaks.

The methods described in this paper allow other types of inferences. For instances, if in a given year the number of cases is far too high to be explained by dengue mechanisms explained above, one can suspect that another virus was introduced in the population. This can either be another strain of dengue or, as we suspect, a new infection like Zika virus (see the final inflection of figure 9).

**ACKNOWLEDGEMENTS**: This work was partially supported by LIM01 HCFMUSP, Fapesp (2014/26229-7 and 2014/26327-9), CNPq, Dengue Tools under the Seventh Framework Programme of the European Community, grant agreement no. 282589, and MS/ FNS (grant no. 777588/2012).